\documentclass{article}                     
\usepackage{graphicx}
\usepackage{amsmath,amssymb}
\usepackage{amscd,float,array,bm}
\usepackage{lscape}
\usepackage{xcolor}
\usepackage{verbatim}

\usepackage{braket}
\usepackage{fancyhdr}
\newcommand{\Ree}{{\rm Re}}
\newcommand{\Ha}{{\rm Ha}}
\newcommand{\Rm}{{\rm Rm}}
\newcommand{\Pm}{{\rm Pm}}

\newcommand{\er}{{\bf\hat e_r}}
\newcommand{\et}{{\bf\hat e_\theta}}

\newcommand{\ez}{{\bf\hat e_z}}
\newcommand{\ve}{{\mathbf{v}}}
\newcommand{\bv}{{\bf B}}
\newcommand{\bvo}{{\bf B_0}}

\newcommand{\rv}{\ensuremath{\mathbf{r}}}

\newcommand{\lp}{\ensuremath{\left(}}
\newcommand{\rp}{\ensuremath{\right)}}

\begin{document}

\title{Modulated rotating waves in the magnetized spherical Couette
  system
}


\author{Ferran Garcia$^{1,2}$, Martin Seilmayer$^{1}$, Andr\'e
  Giesecke$^{1}$ \and Frank Stefani$^{1}$}


\date{%
{\small $^{1}$Helmholtz-Zentrum Dresden-Rossendorf, Bautzner
  Landstra\ss e 400, D-01328 Dresden, Germany\\ $^{2}$Anton Pannekoek
  Institute for Astronomy, University of Amsterdam, Postbus 94249,
  1090 GE Amsterdam, The Netherlands.\\[5.pt]}
    \today
}


  


\maketitle




\begin{abstract}
We present a study devoted to a detailed description of modulated
rotating waves (MRW) in the magnetized spherical Couette system. The
set-up consists of a liquid metal confined between two differentially
rotating spheres and subjected to an axially applied magnetic
field. When the magnetic field strength is varied, several branches of
MRW are obtained by means of three dimensional direct numerical
simulations (DNS). The MRW originate from parent branches of rotating
waves (RW) and are classified according to Rand's~\cite{Ran82} and
Coughling \& Marcus~\cite{CoMa92} theoretical description. We have
found relatively large intervals of multistability of MRW at low
magnetic field, corresponding to the radial jet instability known from
previous studies. However, at larger magnetic field, corresponding to
the return flow regime, the stability intervals of MRW are very narrow
and thus they are unlikely to be found without detailed knowledge of
their bifurcation point. A careful analysis of the spatio-temporal
symmetries of the most energetic modes involved in the different
classes of MRW will allow in the future a comparison with the HEDGEHOG
experiment, a magnetized spherical Couette device hosted at the
Helmholtz-Zentrum Dresden-Rossendorf.\\

\noindent
{\bf Keywords:} magnetohydrodynamics -- nonlinear waves -- bifurcation
theory -- symmetry breaking -- experiments -- astrophysics\\

\noindent
{\bf MSC codes:} 37L15 -- 37L20 -- 65P40 -- 76E25 --
76E30 -- 85A30
\end{abstract}

\section{Introduction}
\label{intro}

The study of how magnetic fields interact with conducting liquids in
rotating spherical containers is crucial for the understanding of many
natural phenomena. For instance, the Earth's~\cite{Jon11} and
Solar~\cite{Rud89} dynamos are driven by global fluid motions. In
addition, the magnetorotational instability (MRI)~\cite{BaHa91} is
considered the best explanation for the transport of angular momentum
in accretion disks around black holes and stars, and also in
protoplanetary disks~\cite{JiBa13}, allowing matter to fall into the
center. Because of its relevance, MRI has been studied experimentally,
with GaInSn between two rotating cylinders at Helmholtz-Zentrum
Dresden-Rossendorf (HZDR)~\cite{SGGRSSH06,SGGHPRS09,SGGGSGRSH14}, and
in Maryland~\cite{SMTHDHAL04} with liquid sodium between
differentially rotating spheres.

In the context of the Maryland sodium experiment~\cite{SMTHDHAL04},
recent numerical~\cite{Hol09,GJG11} and experimental
studies~\cite{KKSS17} have provided alternative interpretations in
terms of typical instabilities in magnetized spherical Couette (MSC)
flows encouraging further research into this problem.  Consider an
electrically conducting liquid confined between two differentially
rotating spheres and subjected to a magnetic field.  Although the
spherical Couette (SC) system has a simple formulation it reveals
immense complexity even without considering the magnetic field.  The
problem is described in terms of the three dimensional incompressible
Navier-Stokes equations with enforced differential rotation
  between the boundaries allowing the development of thin shear
layers (Stewartson layer~\cite{Ste66}) parallel to the rotation axis
along the tangent cylinder.  Moreover, thin Ekman or Ekman-Hartmann
boundary layers~\cite{DoSo07b} appear when the no-slip condition, used
to model planetary dynamos and for comparison with laboratory
experiments, is imposed at the boundaries. These circumstances make
the numerical treatment extremely challenging because of the higher
spatial resolution, even in the study of laminar flows at small
differential rotation rates.

In case of a resting outer sphere (which is the focus of this paper)
three parameters determine the MSC problem: the aspect ratio
$\chi=r_i/r_o$, where $r_i$ ($r_o$) is the radius of the inner (outer)
spherical boundary, the Reynolds number $\Ree$, measuring the strength
of differential rotation, and the Hartmann number $\Ha$ measuring the
intensity of the applied magnetic field. Without magnetic field the
basic axisymmetric SC flow is stable~\cite{Sch86} up to a certain
critical value of $\Ree_c$. Beyond this critical point a
nonaxisymmetric instability develops and a branch of stable or
unstable solutions bifurcates and extends for larger values of
$\Ree$~\cite{MaTu95}. The flow topology of the instability depends
strongly on the gap width. For narrow gaps Taylor-G\"ortler vortices
are formed~\cite{Zik96,Yua12} whereas for sufficiently wide gaps the
instability occurs in form of spiral
waves~\cite{WER99,HJE06,AYS18}. With further increasing $\Ree$,
several bifurcations take place, giving rise to chaotic~\cite{WER99}
and eventually turbulent flows. A comprehensive overview of the
different flow regimes in the SC system with positive or negative
differential rotation was recently given in~\cite{Wic14}.

In contrast, the impact of magnetic fields on a spherical Couette
system has attracted less interest. Although the first instabilities
of the basic axisymmetric MSC flow have been characterised in the
$(\chi,\Ree,\Ha)$ parameter space~\cite{Hol09,TEO11}, most nonlinear
studies briefly explore the parameter space with direct numerical
simulations (DNS) and are focused on addressing the influence of input
physics such as considering different types of boundary conditions for
the magnetic field~\cite{HoSk01,GJG11} (insulating or conducting inner
sphere allowing magnetic lines to pass) or different topologies of the
applied magnetic field (dipolar~\cite{GJG11,FSNS13},
axial~\cite{SMTHDHAL04,Hol09,Kap14}, or a combination of
both~\cite{WeHo10}). Very recently, an exhaustive numerical
study~\cite{KNS18}, related to the Derviche Torneur sodium experiment
in Grenoble~\cite{Bri_et_al11}, revealed several dynamical regimes
where coherent structures coexist with turbulent flows when a dipolar
magnetic field is imposed.

In case of axially applied magnetic fields and small $\Ree$, the
studies~\cite{HoSk01,Hol09,TEO11,Kap14} have shown that the MSC basic
state is equatorially symmetric and stable for all $\Ha$. It is
described as a strong azimuthal flow associated with a meridional
recirculation. At some critical value $\Ree_{\text{c}}$, the basic
flow becomes unstable to non-axisymmetric perturbations. At low $\Ha$
the instability takes the form of an equatorially antisymmetric radial
jet instability at the equatorial plane, whereas at large $\Ha$ the
instability is equatorially symmetric and related to a shear layer at
the tangent cylinder~\cite{HoSk01,Hol09}.  For moderate values of
$\Ha$, between the radial jet and the shear layer instability, it
takes the form of a meridional return flow
instability~\cite{Hol09}. We note that the radial jet and return flow
instabilities are separated by a $\Ha$ interval in which the basic
flow stabilises again~\cite{Hol09,TEO11}. These instabilities have
been recently identified in the HEDGEHOG (\underline{H}ydromagnetic
\underline{E}xperiment with \underline{D}ifferentially
\underline{G}yrating sph\underline{E}res \underline{HO}lding
\underline{G}aInSn) laboratory experiment~\cite{KKSS17} at
Helmholtz-Zentrum Dresden-Rossendorf when increasing the magnetic
field strength ($\Ha$) for a fixed rotation rate ($\Ree$).

Both the SC and the MSC are {\bf{SO}}$(2)\times${\bf{Z}}$_2$
equivariant systems, i.\,e., invariant by azimuthal rotations and
reflections with respect to the equatorial plane, because of the
symmetry of the geometry and the boundary conditions. In
{\bf{SO}}$(2)$ symmetric systems, branches of periodic rotating
waves (RW) appear after primary Hopf
bifurcations~\cite{CrKn91,EZK92,GoSt03} when the axisymmetric
azimuthal symmetry of the basic state is broken. Secondary Hopf
bifurcations~\cite{Ran82,GLM00,GoSt03} give rise to 2-frequency
quasiperiodic amplitude or shape modulated rotating waves (MRW) which
may have different types of spatio-temporal symmetries. A theoretical
characterisation of MRW was first performed in~\cite{Ran82} and
extended in~\cite{GLM00} following several application examples such
as the cylindrical Taylor-Couette (TC) system. Although the MSC
problem is somewhat different to the TC system, it still retains
  the rotational symmetry, and thus the mathematical analysis of MRW
performed in~\cite{CoMa92}, in terms of Floquet theory, also
applies. An algorithm for the identification of MRW in {\bf{O}}(2)
systems was developed in~\cite{Pal02} in the framework of center
bundle construction~\cite{Kru90,GLM00}. In case of the MSC problem
the existence of RW and MRW has been confirmed by experimental
studies~\cite{SABCGJN08} and by DNS~\cite{HoSk01,Hol09,GJG11} but
their dependence upon parameters, especially the Hartmann number, is
still not well understood.

Recently, a continuation method was applied~\cite{GaSt18} to build up
accurate bifurcation diagrams for RW to address the magnetic field
strength dependence. The Floquet analysis of periodic orbits was used
to determine regions of stability of different azimuthal wave numbers
$m$ and to characterise the symmetry of the bifurcated MRW. The
present study continues the work of~\cite{GaSt18} by describing these
MRW in terms of the established theory, using the analysis
of~\cite{Ran82} and~\cite{CoMa92}, and study their dependence upon the
magnetic field strength. Our main focus is to determine the dominant
azimuthal modes involved in each solution and obtain their time scales
in correspondence with the azimuthal symmetries of the flow that can
be measured in the HEDGEHOG experiment. Our analysis is devoted to a
moderate Reynolds number regime and extends previous numerical
studies~\cite{HoSk01,Hol09,TEO11,GaSt18} by determining the regions of
stability of oscillatory waves and describe them in terms of
bifurcation theory. Yet, we leave for future research the study of
chaotic attractors that may originate from the flows described in this
paper at very weak magnetic fields.

The paper is organized as follows: In \S\ \ref{sec:mod} we introduce
the formulation of the problem, the numerical method used to integrate
the model equations, and some background summarising the theory of MRW
classification. In \S\ \ref{sec:bif} the bifurcation diagrams as a
function of $\Ha$ and the different regions of stability of MRW are
presented. The classification of MRW is discussed in
\S\ \ref{sec:clas_mrw} while the analysis of the dominant azimuthal
modes is presented in \S\ \ref{sec:m_dom}.  Finally, in
\S\ \ref{sec:sum} the paper closes with a summary on the results
obtained.

\section{The HEDGEHOG model}
\label{sec:mod}

\subsection{The equations and the numerical method}
\label{sec:eq_num}

We consider a conducting fluid of constant density $\rho$, the
kinematic viscosity $\nu$, magnetic diffusivity $\eta$ and electrical
conductivity $\sigma=1/(\eta\mu_0)$, where $\mu_0$ is the free-space
value for the magnetic permeability. The fluid is confined in a
spherical shell defined by inner and outer radii $r_i$ and $r_o$. The
outer sphere is at rest while the inner rotates at a constant angular
velocity $\Omega$ around the vertical axis $\ez$.

\begin{figure*}[h!]
\begin{center}
  \includegraphics[scale=1.]{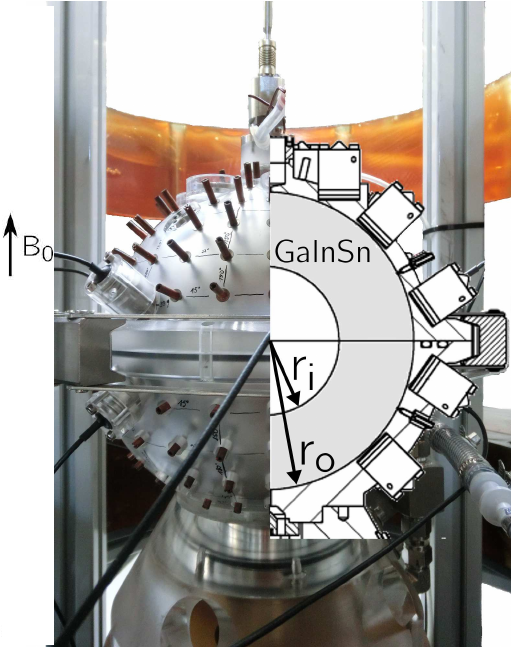}
\end{center}  
\caption{HEDGEHOG experimental configuration. Detail of the rotating
  spheres showing the ultrasonic Doppler velocimeter (UDV) sensors
  (thick cylinders) and the electric potential probes (thin needles)
  attached to the outer sphere.}
\label{fig:hed_exp}   
\end{figure*}

To compare with the HEDGEHOG laboratory experiment~\cite{KKSS17} the
flow is subjected to a uniform axial magnetic field $\bvo=B_0
\cos(\theta)\er-B_0 \sin(\theta)\et$, $\theta$ being the colatitude
and $B_0$ the magnetic field strength (Fig.~\ref{fig:hed_exp}). To
obtain the dimensionless governing equations the length, time,
velocity and magnetic field are scaled using the characteristic
quantities $d=r_o-r_i$, $d^2/\nu$, $r_i\Omega$ and $B_0$,
respectively. We adopt the inductionless approximation $\Rm\ll 1$
valid in the limit of low magnetic Reynolds number $\Rm=\Omega r_i
d/\eta$. This is reasonable in the case of the HEDGEHOG experiment
because the fluid has very low magnetic Prandtl number
$\Pm=\nu/\eta\sim O(10^{-6})$ (eutectic alloy GaInSn~\cite{MBCN08})
and only moderate Reynolds numbers $\Ree=\Omega r_i d/\nu\sim 10^3$
are considered giving rise to $\Rm=\Pm\Ree \sim 10^{-3}$.

The magnetic field is decomposed as $\bv=\ez+\Rm{\bf b}$. By
neglecting terms $O(\Rm)$, the Navier-Stokes and induction equations
become
\begin{align}
  \partial_t\ve+\Ree\lp\ve\cdot\nabla\rp\ve &=
-\nabla p+\nabla^2\ve+\Ha^2(\nabla\times {\bf b})\times\ez, \label{eq:mom_less}   \\
 0& = \nabla\times(\ve\times\ez)+\nabla^2{\bf b}, \label{eq:ind_less}\\
\nabla\cdot\ve=0, &\quad \nabla\cdot{\bf b}=0.\label{eq:div}
\end{align}
In this inductionless approximation the MSC system depends upon three
non-dimensional numbers: the Reynolds number, the Hartmann number and
the aspect ratio
\begin{equation*}
  \Ree=\frac{\Omega r_i d}{\nu}, \quad
  \Ha=\frac{B_0d}{\sqrt{\mu_0\rho\nu\eta}}=B_0d\sqrt{\frac{\sigma}{\rho\nu}},\quad
  \chi=\frac{r_i}{r_o}.
\end{equation*}
No-slip ($v_r=v_\theta=v_\varphi=0$) at $r=r_o$ and constant rotation
($v_r=v_\theta=0,~v_\varphi= \sin{\theta}$) at $r=r_i$ are the
boundary conditions imposed on the velocity field. For the
magnetic field, insulating exterior regions are considered in
accordance with the experimental setting, see~\cite{HoSk01} for more
details.

The equations are discretized and integrated with the same method as
described in~\cite{GNGS10} and references therein. The velocity field
is expressed in terms of toroidal, $\Psi$, and poloidal,
$\Phi$, potentials
\begin{equation}
  \ve=\nabla\times\lp\Psi\rv\rp+\nabla\times\nabla\times\lp\Phi\rv\rp,
\label{eq:pot}  
\end{equation}
which are expanded in spherical harmonics in the angular coordinates
($\rv=r~\er$ is the position vector). In the radial direction a
collocation method on a Gauss--Lobatto mesh of $N_r$ points is
used. Specifically, the solution vector $u=(\Psi,\Phi)$
(Eq.~\ref{eq:pot}) is expanded in spherical harmonic series up to
degree $L_{\text{max}}$
\begin{eqnarray}
  \Psi(t,r,\theta,\varphi)=\sum_{l=0}^{L_{\text{max}}}\sum_{m=-l}^{l}{\Psi_{l}^{m}(r,t)Y_l^{m}(\theta,\varphi)},\label{eq:serie_psi}\\
  \Phi(t,r,\theta,\varphi)=\sum_{l=0}^{L_{\text{max}}}\sum_{m=-l}^{l}{\Phi_{l}^{m}(r,t)Y_l^{m}(\theta,\varphi)},\label{eq:serie_phi}
\end{eqnarray}
with $\Psi_l^{-m}=\overline{\Psi_l^{m}}$,
$\Phi_l^{-m}=\overline{\Phi_l^{m}}$, $\Psi_0^0=\Phi_0^0=0$ to uniquely
determine the two scalar potentials, and
$Y_l^{m}(\theta,\varphi)=P_l^m(\cos\theta) e^{im\varphi}$, $P_l^m$
being the normalized associated Legendre functions of degree $l$ and
order $m$. The code is parallelized in the spectral and in the
physical space by using OpenMP directives. We use optimized libraries
(FFTW3~\cite{FrJo05}) for the FFTs in $\varphi$ and matrix-matrix
products (dgemm GOTO~\cite{GoGe08}) for the Legendre transforms in
$\theta$ when computing the nonlinear terms.

For the time integration, high order implicit-explicit backward
differentiation formulas (IMEX--BDF)~\cite{GNGS10} are used. In the
IMEX method we treat the nonlinear terms explicitly in order to avoid
solving nonlinear equations at each time step. The Lorenz term is also
treated explicitly, which may necessitate a reduced time step in
comparison with an implicit treatment. However, this is not a serious
issue when moderate $\Ha$ are considered, as is the case for the
present study.  The use of \textit{matrix-free} Krylov methods
(GMRES~\cite{SaSc86} in our case) for the linear systems facilitates
the implementation of a suitable order and time stepsize control for
the time integration (see~\cite{GNGS10} for details on the
implementation).

\subsection{Background for Modulated Rotating Waves}
\label{sec:eq_waves}


The discretization of the MSC equations
Eqs.~(\ref{eq:mom_less}-\ref{eq:div}) leads to a system of
$n=(2L_{\text{max}}^2+4L_{\text{max}})(N_r-1)$ ordinary differential
equations (ODE) of the form
\begin{align}
L_0\partial_tu=L u+B(u,u),
\label{eq:op_eq}
\end{align}
where $L_0$ and $L$ are linear operators which include the boundary
conditions (see~\cite{GNGS10} for details). The first operator is
invertible and the operator $L$ depends on $\Ha$ (the control
parameter of the present study) and includes all the linear terms,
whereas the non-linear (quadratic) terms are included in the bilinear
operator $B$.

\begin{figure}
  \hspace{2.mm} \includegraphics[scale=0.83]{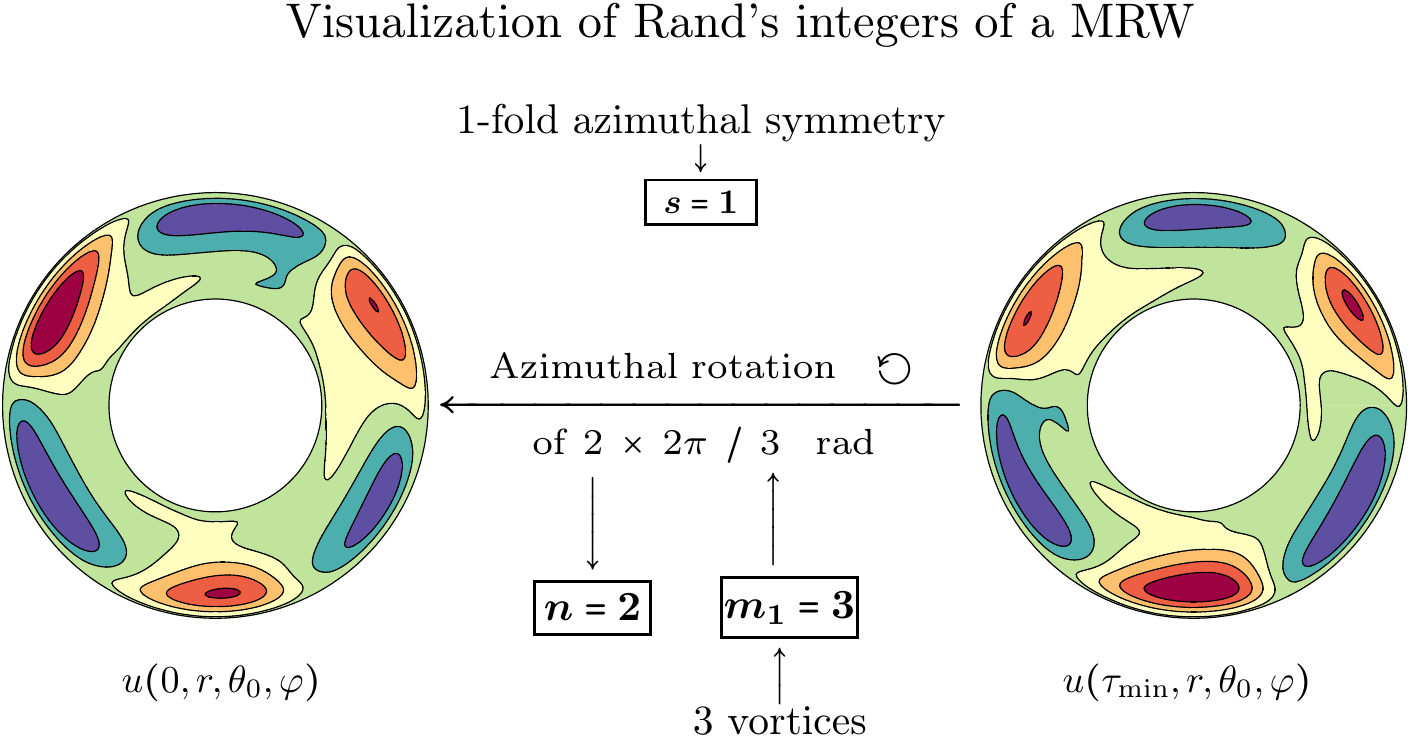}
\caption{Contour plots of the nonaxisymmetric ($m>0$) radial velocity
  on colatitudinal cross-sections at $t=0$ and $t=\tau_{\text{min}}$
  for a MRW at $\Ha=3.5$. The main azimuthal drift corresponding to
  $\omega$ has been removed. The position of colatitudinal cuts is
  one degree above the equator.}
\label{fig:mrw_sch}   
\end{figure}

The system is {\bf{SO}}$(2)\times${\bf{Z}}$_2$-equivariant,
{\bf{SO}}$(2)$ generated by azimuthal rotations, and {\bf{Z}}$_2$ by
reflections with respect to the equatorial plane. For fixed $\Ree$, at
a critical $\Ha_{\text{c}}$ the basic axisymmetric ($m=0$) flow is
unstable to nonaxisymmetric perturbations~\cite{Hol09,TEO11} giving
rise to a branch of RWs~\cite{EZK92}. These are solutions in which a
fixed flow pattern with $m_1$-fold azimuthal symmetry is rotating at a
frequency $\omega$ in the azimuthal direction~\cite{Ran82}. For the
MSC system, RW have been recently computed~\cite{GaSt18}, as a function
of $\Ha$, by means of continuation methods for periodic
orbits~\cite{SaNe16}. However, RW can also be computed as fixed
points~\cite{SGN13,TLW19} of the MSC system written in a reference
frame rotating with the same frequency $\omega$:
\begin{equation}
L_0\partial_tu= L(p)u+B(u,u) + \omega L_0\partial_{\varphi} u.
\label{eq:wave_eq}
\end{equation}
This is because RW satisfy $u(t,r,\theta,\varphi-\omega t)=\tilde
u(r,\theta,\tilde\varphi)$, with $\tilde\varphi=\varphi-\omega
t$.
A branch of RW, rotating at a frequency $\omega$ and with $m_1$-fold
azimuthal symmetry, undergoes secondary Hopf bifurcations at a certain
critical value of $\Ha$ giving rise to a branch of MRW. According
to~\cite{Ran82} these are $\tau$-periodic solutions of
Eq.~(\ref{eq:wave_eq}) (i.\,e. in a reference frame rotating with
frequency $\omega$) for which there exist a basic (minimal) time
$\tau_{\text{min}}>0$ and an integer $0\le n<m_1/s$ such that
\begin{equation}
u(t,r,\theta,\varphi)=u(t+\tau_{\text{min}},r,\theta,\varphi +2\pi n/m_1) \quad
\forall t,~\forall \varphi.
\label{eq:mod_trav}
\end{equation}
The spatio-temporal symmetry of each MRW is then described with three
integers $(m_1,n,s)$. They are the $m_1$-fold azimuthal symmetry of
the parent RW, the integer $n$ related with the minimal period
$\tau_{\text{min}}$ in Eq.~(\ref{eq:mod_trav}), and the $s$-fold
azimuthal symmetry of the MRW (see the scheme in
  Fig.~\ref{fig:mrw_sch}). We note that $s$ can be deduced a priori
from the stability analysis (Floquet theory) of a RW close to the
bifurcation point~\cite{CoMa92}. In this way, if at the bifurcation
point the dominant Floquet multiplier has $m_2$-fold azimuthal
symmetry then the azimuthal symmetry of the bifurcated MRW will be
$s=\text{GCF}(m_1,m_2)$ (GCF stands for great common factor). In the
original system (Eq.~(\ref{eq:op_eq})) MRW are quasiperiodic and any
frequency obtained from a time spectrum analysis will be a linear
combination (with integer coefficients) of Rand's fundamental
frequencies~\cite{Ran82}
\begin{equation}
  \Omega_M=2\pi/\tau_{\text{min}}, \quad \Omega_W=s(\omega+n\Omega_M)/m_1.
\label{eq:freq_rand}  
\end{equation}

We note that the MRW period $\tau$ seen when integrating system
Eq.~(\ref{eq:wave_eq}) (i.\,e. rotating with frequency $\omega$) may
not be minimal (this only happens when $n=0$). However, there exists a
frequency $\omega_{\text{min}}$ and its associated system
(Eq.~(\ref{eq:wave_eq})) for which the MRW will exhibit
$\tau_{\text{min}}$ periodicity. Detecting this minimal period is
important to efficiently obtain MRW with continuation methods for
periodic orbits~\cite{GNS16}, as these methods require a large number
of time integrations over one modulation period. We note that
azimuthally averaged properties of a MRW (either obtained with
Eq.~(\ref{eq:op_eq}) or Eq.~(\ref{eq:wave_eq})) will naturally reflect
the minimal period as any azimuthal drift is removed~\cite{Pal02}.

For a MRW close to the bifurcation point its rotating frequency is
close to that of the parent RW. An approximation for the second
frequency (related with the modulation period) can be obtained by
means of Floquet theory in the way we now describe. Let
$\mu\in\mathbb{C}$ be a Floquet multiplier and $\lambda\in\mathbb{C}$
be a Floquet exponent~\cite{JoSm07} of a RW of the MSC system
(Eq.~(\ref{eq:op_eq})), rotating at a frequency $\omega$ and with
$m_1$-fold azimuthal symmetry. The period of the RW is
$T=2\pi/m_1\omega$. The Floquet multipliers and exponents are related
by $\mu=e^{T\lambda}$. At a Hopf bifurcation point a branch of MRW is
born and a modulation period can be approximated by
$\tau_{\text{b}}=2\pi/\Im(\lambda)$. At the bifurcation point
$\Re(\lambda)=0$ and thus $T \Im(\lambda)=\text{Arg}(\mu)$ with
$\text{Arg}(\mu)$ the complex argument
($\mu=|\mu|e^{i\text{Arg}(\mu)}$). Then we have
\begin{equation}
\tau_{\text{b}}=2\pi T/\text{Arg}(\mu)= 4\pi^2 /m_1\omega\text{Arg}(\mu).
\label{eq:taub}
\end{equation}
We note that $\tau_{\text{b}}$ may neither be the minimal modulation
period $\tau_{\text{min}}$ nor the modulation period $\tau$ observed
when integrating the system Eq.~(\ref{eq:wave_eq}). Nevertheless,
$\omega$ and $2\pi/\tau_{\text{b}}$ are fundamental frequencies as
well, so they can be expressed as functions of $\Omega_M$ and
$\Omega_W$~\cite{Ran82}. This is useful because Floquet multipliers of
RW at bifurcation points have been already obtained in~\cite{GaSt18}
and thus can be used to check the frequencies obtained from a time
series analysis of DNS of MRW close to the corresponding bifurcation
points.

Besides Rand's classification of MRW~\cite{Ran82} the authors
of~\cite{CoMa92} provide an equivalent classification by deriving a
functional form for MRW, within the context of Taylor-Couette flows,
in terms of bifurcation and Floquet theory. According to~\cite{CoMa92}
a MRW may be described by $(m_1,c_1,m_2,c_2)$ with $m_1$ and
$c_1=\omega$ being the azimuthal symmetry and rotating frequency
of the underlying RW, and $m_2$ and $c_2=c_1+\omega_M/m_2$
($\omega_M=2\pi/\tau$), the azimuthal symmetry and frequency of the
Floquet mode, respectively, associated to the modulation (with period
$\tau$) seen in the frame rotating with frequency $c_1$.

\section{Bifurcation diagrams}
\label{sec:bif}


For fixed $\chi=0.5$ and in the absence of magnetic field the basic
axisymmetric flow becomes unstable at sufficiently large
$\Ree_{\text{c}}\approx 489$ (see~\cite{HJE06}). Then, for $\Ree=10^3$
the flow is nonaxisymmetric taking the form of a radial jet
instability~\cite{HJE06}. The linear stability analysis of the basic
state of~\cite{Hol09,TEO11} showed that with increasing magnetic
forcing ($\Ha$) the basic state is recovered at $\Ha_{\text{c}}=12.2$,
but loses its stability again at $\Ha_{\text{c}}=25.8$ giving rise to
a meridional circulation and a return flow instability. With further
increase of $\Ha$ the basic state becomes stable again at
$\Ha_{\text{c}}=79.4$ taking the form of a strong shear layer parallel
to the rotation axis and close to the inner boundary.

At the same $\Ree=10^3$, branches of unstable/stable rotating waves
with azimuthal symmetry $m=2,3,4$ bifurcated from the basic state at
the critical points have been recently computed in~\cite{GaSt18} for
the three types of instabilities, i.\,e., for the radial jet
instability in the range $\Ha\in[0,12.2]$ and for the return
flow and shear layer instabilities in the range
$\Ha\in[25.8,79.4]$. The stability analysis (Floquet) of these waves
provided the values of the bifurcation points to branches of MRW which
are described in this section.

The code and results for the computation of RW and their stability
analysis were validated in section 3(c) of~\cite{GaSt18}. In the
present study the same spatial resolutions are employed as the MRW are
obtained in the same range of parameters. We use $N_r=40$ radial
collocation points and a spherical harmonics truncation parameter
$L_{\text{max}}=84$ giving rise to a nonlinear system of $563472$
degrees of freedom (DOF). When increasing the resolution to $N_r=60$
and $L_{\text{max}}=126$ ($1903104$ DOF), errors below $1\%$ are
obtained (see table 1 of~\cite{GaSt18}). Numerical experiments with a
VSVO time step method (see~\cite{GNGS10}) allows us to determine an
appropriate fixed time step (of order $\Delta t=5\times 10^{-6}$) for
an accurate fixed time step integration.

We compute MRW using DNS and start each branch with initial conditions
built from a RW and its Floquet multiplier close to the bifurcation
point. Successive MRW on the branches are obtained from those at
slightly different $\Ha$. An alternative procedure allows us to obtain
additional branches of MRW without knowing their parent branch of RW
(the point where they bifurcate). We set as initial condition a
slightly perturbed unstable RW and integrate in time (DNS) until the
perturbations are saturated and an attractor is reached. By means of
time spectrum analysis we infer the quasiperiodic character of the
DNS. By imposing azimuthal symmetry constraints on the DNS (by
retaining only certain modes on the spherical harmonic expansion) we
are able to capture unstable MRW.

\begin{figure*}[h!]
\begin{center}
  \includegraphics[scale=1.8]{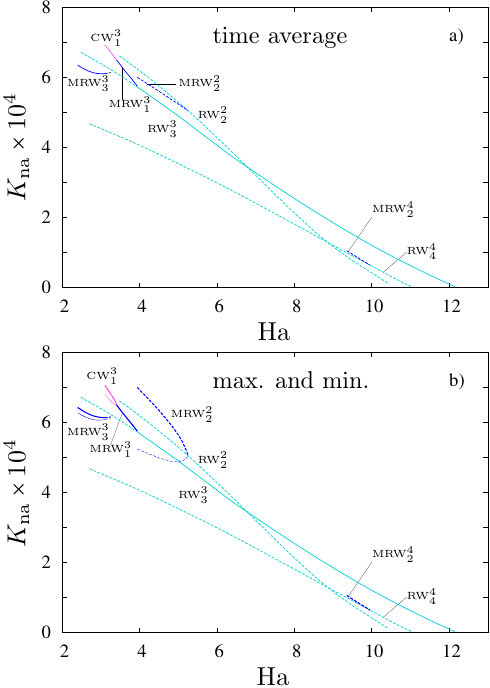}
\end{center}  
\caption{Bifurcation diagrams for rotating and modulated rotating
  waves (RW/MRW) corresponding to the equatorially asymmetric radial
  jet instability. The label notation is RW/MRW$_s^{m_{\text{max}}}$,
  $s$ being the azimuthal symmetry of the waves and
  $m_{\text{max}}\ne0$ their most energetic azimuthal wave
  number. A stable branch of complex waves with 3 frequencies,
    $m_{\text{max}}=3$, and azimuthal symmetry $s=1$ is also included
    (CW$_1^3$). For $\Ha\to 0$ different complex and even chaotic
    flows are identified but their description is out of scope of the
    present study.  The volume averaged nonaxisymmetric ($m>0$)
  kinetic energy $K_{\text{na}}$ is plotted versus the Hartmann number
  $\Ha$. (a) Time average and (b) maximum (thick line) and minimum
  (thin line) of the time series.  Solid/dashed lines mean
  stable/unstable waves.  }
\label{fig:bif_diagr}   
\end{figure*}

Figure~\ref{fig:bif_diagr}(a,b) displays the time average and
minimum/maximum values of the volume averaged nonaxisymmetric ($m>0$)
kinetic energy $K_{na}$ versus the Hartmann number in the low magnetic
field regime corresponding to the radial jet instability. Branches for
RW with $m$-fold azimuthal symmetry $m=2,3,4$ were obtained
in~\cite{GaSt18}, where several bifurcation points along the branches
were computed as well. This information is summarised in
Table~\ref{table:bif_stab}. A branch of RW or MRW is denoted by
RW$_s^{m_{\text{max}}}$ or MRW$_s^{m_{\text{max}}}$ when the waves
belonging to the branch have $s$-fold azimuthal symmetry and their
most energetic azimuthal wave number (with the exception of $m=0$) is
$m_{\text{max}}$. For instance, a wave in the MRW$_1^{3}$ branch has
$1$-fold azimuthal symmetry (i.~e. invariant by $2\pi$ azimuthal
rotations) and $K_3\ge K_m ~(m\ne0)$, $K_m$ being the mean rms kinetic
energy contained in the azimuthal wave number $m$.

The branches of RW$_2^2$ and RW$_4^4$ are unstable and thus the
bifurcated MRW$_2^2$ (at $\Ha_{\text{c}}=5.27$) and MRW$_2^4$ (at
$\Ha_{\text{c}}=9.98$) are unstable as well. Because of the $m=2$
azimuthal symmetry of the corresponding Floquet mode at the
bifurcation, the branches of MRW$_2^2$ and MRW$_2^4$ can be computed
with $m=2$ azimuthal symmetry constraints imposed on the DNS. In
contrast to RW$_2^2$ or RW$_4^4$, the branch of RW$_3^3$ has a stable
$\Ha$ interval. The RW$_3^3$ branch corresponds to the branch of the
most unstable linear mode of the basic state~\cite{TEO11}, emerging at
$\Ha_{\text{c}}=12.2$. At $\Ha_{\text{c}}=3.95$, RW$^3_3$ becomes
unstable by decreasing $\Ha$, at a Hopf bifurcation, giving rise to a
branch of stable MRW$^3_1$. By further decrease of $\Ha$ (along the
RW$^3_3$ branch) a second Hopf bifurcation, which does not break the
$m=3$ azimuthal symmetry, gives rise to a branch of unstable MRW$_3^3$
at $\Ha_{\text{c}}=3.33$. Again these waves can be computed with
azimuthal symmetry constraints as done for the case of MRW$_2^2$ or
MRW$_2^4$. At $\Ha_{\text{c}}\approx 3.1$ the branch of MRW$_3^3$
becomes stable to arbitrary random perturbations with $m=1$-fold
azimuthal symmetry.

\begin{table}[ht]
\caption{Critical parameters of the equatorially asymmetric RWs at the
  bifurcations where they change the stability ($|\mu|=1$). They are
  obtained by inverse interpolation with a polynomial of degree
  $5$. The sign of the real part of the Floquet exponent of the
  two leading eigenfunctions before and after the transition is
  shown. The azimuthal symmetry ($2\pi/m_2$ azimuthal periodicity) of
  the eigenfunction with a change of sign is also stated. We use ES=1
  for equatorially symmetric flows, ES=0 otherwise. The rotating
  frequency $\omega$ and the fundamental frequency of the time
  spectrum $f_{\omega}=m\omega/2\pi$, the argument of the
  Floquet multiplier $\text{Arg}(\mu)$ and the corresponding
  modulation period $\tau_{\text{b}}$ and fundamental frequency
  $f_{\text{b}}=1/\tau_{\text{b}}$ (see Eq.~(\ref{eq:taub})), are also
  tabulated.}
\label{table:bif_stab}
\scalebox{0.86}{
\begin{tabular}{llllllllll}    
\hline\noalign{\smallskip}
 $m$   & ES     & Signs at transition              & $\Ha_{\text{c}}$ & $m_2$ & $\omega$ & $f_{\omega}$& $\text{Arg}(\mu)$ & $\tau_{\text{b}}$  & $f_{\text{b}}$ \\
\noalign{\smallskip}\hline\noalign{\smallskip}
 $2$   &  $0$   & $(+,+)\rightarrow(+,-)$ & $5.27$  & $2$   & $132.40$ & $42.145$  & $0.291$           & $0.512$            & $1.952$      \\
 $3$   &  $0$   & $(+,+)\rightarrow(+,-)$ & $3.33$  & $3$   & $138.88$ & $66.310$  & $2.131$           & $0.044$            & $22.492$     \\
 $3$   &  $0$   & $(+,-)\rightarrow(-,-)$ & $3.95$  & $1$   & $139.07$ & $66.399$  & $0.774$           & $0.122$            & $8.181$      \\
 $4$   &  $0$   & $(+,+)\rightarrow(+,-)$ & $9.98$  & $2$   & $136.08$ & $86.630$  & $3.131$           & $0.023$            & $43.173$     \\
\noalign{\smallskip}\hline\noalign{\smallskip}                                   
 $2$   &  $1$   & $(+,+)\rightarrow(+,-)$ & $29.98$ & $2$   &  $56.44$ & $17.966$  & $0.632$           & $0.553$            & $1.807$      \\
 $4$   &  $1$   & $(-,-)\rightarrow(+,-)$ & $31.95$ & $1$   & $101.62$ & $64.691$  & $2.076$           & $0.047$            & $21.377$     \\
\noalign{\smallskip}\hline
\end{tabular}}
\end{table}

\begin{figure*}
  \includegraphics[scale=0.85]{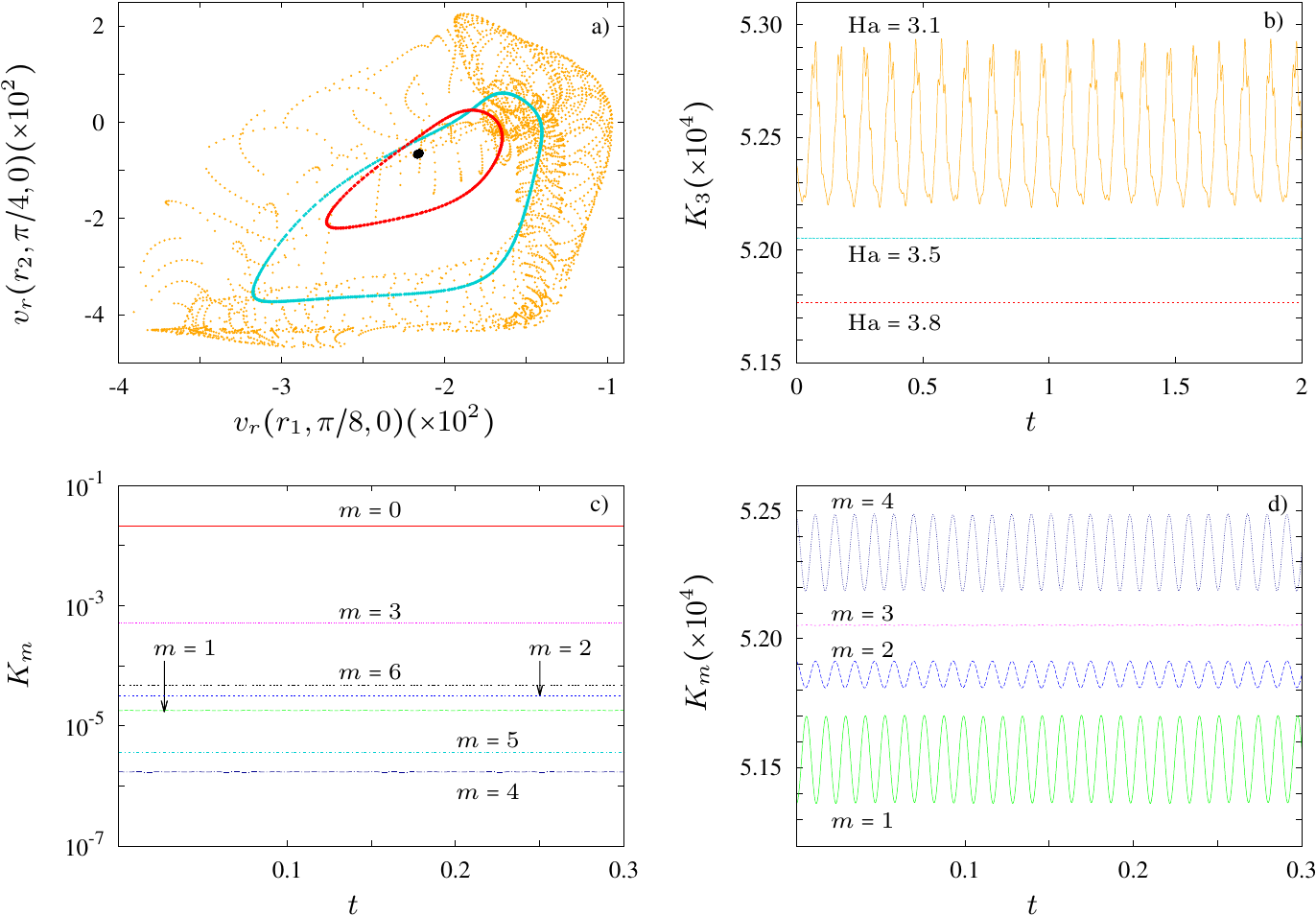}
\caption{(a) Poincar\'e sections defined by $v_r(r_2,\pi/8,0)=-6\times
  10^{-2}$ for a RW$_3^3$ at $\Ha=4$ (black point) for MRW$_1^3$ at
  $\Ha=3.8,3.5$, from smaller (red) to larger (blue) closed curves,
  and a CW$_1^3$ at $\Ha=3.1$ (cloud of orange dots). The radial
  positions are $r_1=r_i+0.15d$ and $r_2=r_i+0.5d$. (b) The
  oscillations of the $m=3$ component of the flow are noticeable from
  the 3rd instability (the onset of 3rd frequency via Hopf
  bifurcation). From top to bottom $K_3$ for $\Ha=3.1$ (orange),
  $\Ha=3.5$ (blue), and $\Ha=3.8$ (red). (c) Time series of kinetic
  energy densities $K_m$ for a MRW$_1^3$ at $\Ha=3.5$. (d) Detail of
  (c) with scaled $K_m$ to stress the small relative variance of the
  oscillations. The scaling factors are $28.7,16.4,1,306$ for
  $m=1,2,3,4$, respectively.}
\label{fig:ts_m3} 
\end{figure*}

\begin{figure*}
  \begin{center}
    \includegraphics[scale=1.6]{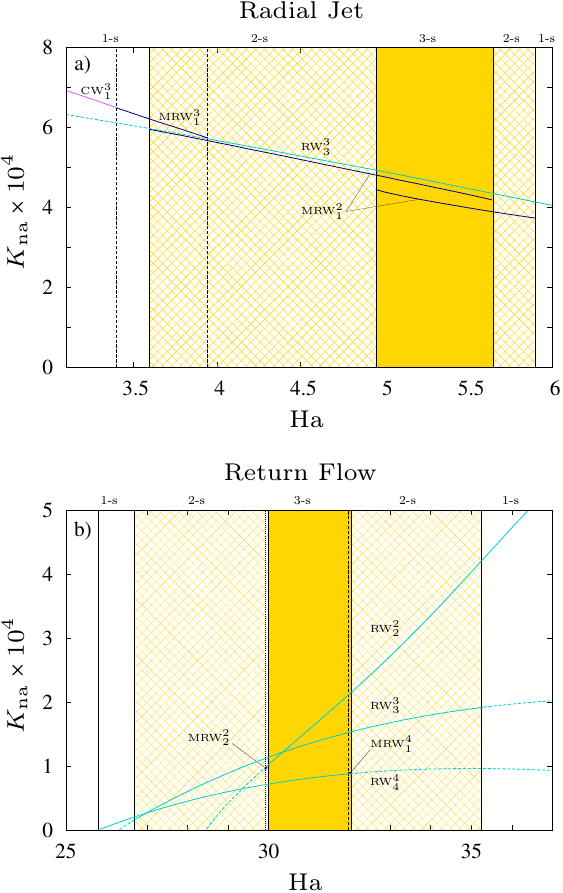}
  \end{center}    
\caption{Bifurcation diagrams of the volume averaged nonaxisymmetric
  ($m>0$) kinetic energy $K_{\text{na}}$ versus the Hartmann number
  $\Ha$. (a) equatorially asymmetric radial jet instability and (b)
  equatorially symmetric return flow instability. Notation is as in
  Fig.~\ref{fig:bif_diagr}. The different regions of multistability
  are highlighted with bands: one stable solution (empty),
  bi-stability (cross-hatched) and tri-stability
  (solid). Vertical dashed lines separate different type of solutions
  in each band. In contrast to the radial jet instability, less
  different types of modulated rotating waves, on very narrow regions,
  have been found for the return flow instability (panel (b)). Only
  MRW$_1^4$ are stable. The dotted vertical line delimit the thin band
  where unstable MRW$_2^2$ are found.}
\label{fig:bif_stab}   
\end{figure*}

By taking initial conditions on the MRW$_1^3$ branch, a stable branch
of quasiperiodic waves, with three frequencies and with
$m_{\text{max}}=3$ and azimuthal symmetry $m=1$ (labelled CW$_1^3$),
is found if $\Ha$ is decreased beyond the threshold of
$\Ha_{\text{c}}=3.4$. This sequence of bifurcations --basic steady
state, RW$_3^3$, quasiperiodic with two frequencies MRW$_1^3$ and
quasiperiodic with three frequencies MRW$_1^3$ -- seen at low $\Ha$
for the radial jet instability with $m_{\text{max}}=3$ corresponds to
the Ruelle-Takens scenario~\cite{RuTa71,Eck81} which is typical for
systems with symmetry~\cite{Ran82,GLM00}. Figure~\ref{fig:ts_m3}(a)
displays Poincar\'e sections (see figure caption) for a selection of
four different flows which evidence the sequence of bifurcations. At
$\Ha=4$ the Poincar\'e section of a RW$_3^3$ (periodic orbit) is a
point while at the lower values $\Ha=3.8$ (red curve) and $\Ha=3.5$
(blue curve) the sections of MRW$_1^3$ (invariant tori) are closed
curves. At $\Ha=3.1$ the cloudy Poincar\'e section exhibits features
of a 3 frequency solution. We note that an accurate computation of
bifurcation points to MRW with 3 frequencies in {\bf{SO}}$(2)$ systems
could be done efficiently as in~\cite{GNS16} by considering MRW as
periodic orbits.

An interesting property of MRW$_1^3$, but also of MRW$_2^4$, is that
the amplitude of oscillations of azimuthally averaged properties
($K_{na}$ in particular) is very small. In fact, the maximum and
minimum curves of Fig.~\ref{fig:bif_diagr}(b) are indistinguishable
from each other. This may lead to confuse MRW$_1^3$ or MRW$_2^4$ with
RW as their azimuthally averaged properties appear to be nearly
constant. This is illustrated in Fig.~\ref{fig:ts_m3}(b) displaying
the kinetic energy contained in the $m=3$ azimuthal wave number
($K_3$) for three different flows. For solutions lying on the
MRW$_1^3$ branch ($\Ha=3.5,3.8$) the kinetic energy $K_3$ seems to be
constant contrasting to the oscillatory behaviour seen at
$\Ha=3.1$. The amplitudes of oscillation in the kinetic energy wave
number $m$ spectra for the MRW$_1^3$ at $\Ha=3.5$ shown in
Fig.~\ref{fig:ts_m3}(c) are extremely small and become only visible
when the time series of $K_m$ are examined in detail
(Fig.~\ref{fig:ts_m3}(d)).

Aside the branches of MRW shown in Fig.~\ref{fig:bif_diagr} and born
at the bifurcation points summarised in Table~\ref{table:bif_stab}, we
have found two additional branches with $m=1$-fold azimuthal symmetry
and $m_{\text{max}}=2$ labelled as MRW$_1^2$. These MRW correspond to
the radial jet instability and display noticeable oscillations of the
azimuthally averaged physical properties. They are stable along all
the branch as no azimuthal constraint is imposed on the DNS. The
branches of MRW$_1^2$ are displayed in Fig.~\ref{fig:bif_stab}(a)
together with the branches of RW$_3^3$, MRW$_1^3$ and CW$_1^3$ already
shown in Fig.~\ref{fig:bif_diagr}(a). The intervals of stability of
the different waves overlap, giving rise to regions of multistability
of two and even three different types of waves. In the limits of the
3-stability region we have found hysteretic behaviour between the two
branches of MRW$_1^2$. Because we are using DNS the branches of
MRW$_1^2$ are lost close to $\Ha=3.5$ (or $\Ha=6$) as they
  become unstable and all the azimuthal symmetries are broken. The
use of continuation methods~\cite{SaNe16,GNS16} may help to understand
their origin.

According to Fig.~\ref{fig:bif_stab}(a) for the radial jet instability
(at low $\Ha$) oscillatory solutions (MRW and 3 frequency waves) are
stable in a relatively wide interval of Hartmann numbers. This
contrasts with the dynamical behaviour obtained in the return flow
regime (moderate $\Ha$) for which very narrow intervals of stability
of only two types of MRW are found (see Fig.~\ref{fig:bif_stab}(b) and
Table~\ref{table:bif_stab}). By increasing $\Ha$ further, only RW are
found for the case of the shear layer instability~\cite{GaSt18} at the
selected $\Ree=10^3$ of the present study. According to these results,
increasing magnetic field results in a decrease of the flow complexity
in the sense that less different types of waves are obtained and the
flow solutions exhibit a simpler time dependence. For example, for
$\Ha\in[35.24,79.42]$ only RW with $m=2$-fold azimuthal symmetry are
found~\cite{GaSt18}.

\section{Classification of MRW}
\label{sec:clas_mrw}

Several examples of MRW (belonging to each of the branches previously
described) are classified in this section following
Rand's~\cite{Ran82} and Coughling \& Marcus~\cite{CoMa92} theoretical
work, previously summarised in Sec.~\ref{sec:eq_waves}. The
  numerical approach for the classification and the flow patterns of
  the MRW are described as well.

\subsection{Classification algorithm}

Our procedure is based on DNS (after filtering the initial transient)
of the MSC system equations (Eq.~(\ref{eq:op_eq})) as well as on a
rotating frame with selected rotating frequency
(Eq.~(\ref{eq:wave_eq})) in the way we now describe:
\begin{itemize}
\item[1] Given an initial condition a time evolution of system
  Eq.~(\ref{eq:op_eq}) is performed to determine $f_{\omega}$. The
  latter corresponds to the main peak of the frequency spectrum and
  can be found accurately by using Laskar's algorithm~\cite{Las93}
  applied to a time series of the radial velocity component $v_r(t,p)$
  picked up at a point inside the shell
  $p=(r,\theta,\varphi)=((r_i+r_o)/2,\pi/4,0)$. The azimuthal symmetry
  $s$ of the MRW and the azimuthal symmetry $m_1$ of the underlying
  rotating wave are detected by computing the azimuthal wave number's
  kinetic energy spectra. We have found in all cases that
  $m_1=m_{\text{max}}$ corresponds to the wave number of maximum
  amplitude (except $m=0$). The rotating frequency of the MRW is
  $\omega=2\pi f_{\omega}/m_1$.
\item[2] Once $\omega$ is known, we integrate system
  Eq.~(\ref{eq:wave_eq}) approximately one modulation period $\tau$ (by
  imposing $|v_r(0,p)-v_r(\tau,p)|/v_r(0,p)<10^{-3}$) and capture the
  flow patterns at times $t=0,\tau/4,\tau/3,\tau/2$. If $\tau$ is not
  the minimal period then the patterns at time $t=0$ and at time
  $t=\tau/j$, for some $1<j\le m_1$, should differ by an azimuthal
  rotation of $2\pi n/m_1$ degrees. Then, the minimal rotation period
  $\tau_{\text{min}}$ and the integer $n$ appearing in the labels
  $(m_1,n,s)$ of Rand's classification~\cite{Ran82} are
  identified. Fundamental Rand's frequencies are easily obtained from
  Eq.~(\ref{eq:freq_rand}).
\item[3] Aside $c_1=\omega$ and $m_1$ the classification
  of~\cite{CoMa92} involves the integer $m_2$ and frequency
  $c_2=c_1+\omega_M/m_2$ with $\omega_M=2\pi/\tau$. The integer $m_2$
  corresponds to the azimuthal symmetry of the underlying Floquet mode
  which can be inferred from the DNS by removing the multiples of
  $m_1$ in the spherical harmonics expansion.
\end{itemize}
\begin{table}[ht] 
  \caption{Rand's classification~\cite{Ran82} for the modulated
    rotating waves: the integers $m_1$, $n$ and $s$ and the
    frequencies $f_{\omega}=m_1\omega/2\pi$,
    $f_M=1/\tau_{\text{min}}$, $f_W=s(f_{\omega}/m_1+nf_M)/m_1$,
    $f_{\tau}=1/\tau$ and
    $f_{\omega_{\text{min}}}=m_1\omega_{\text{min}}/2\pi$.  The
    modulation period $\tau$ is that exhibited by any scalar field, at
    a point inside the shell, when integrating the system rotating at
    a frequency $\omega$ (i.\,e. removing the main azimuthal drift). The
    modulation period $\tau_{\text{min}}$ corresponds to that of any
    azimuthally averaged property and that exhibited by any scalar
    field, at a point inside the shell, when integrating the system
    rotating at a frequency $\omega_{\text{min}}$. If
    $\tau=k\tau_{\text{min}}$ for any integer $k>1$ then
    $\omega_{\text{min}}=\omega-2\pi/\tau$.  The classification
    of~\cite{CoMa92} involves $m_1$ and $f_{\omega}$ but also $m_2$
    (the azimuthal symmetry of the Floquet multiplier) and
    $f_2=m_2f_{\omega}/m_1+f_{\tau}$.  For the MRW$_1^2$ at $\Ha=5$,
    (1) stands for branch 1 and (2) for branch 2.}
\label{table:rand}
\begin{center}
\scalebox{0.9}{  
\begin{tabular}{llllllllllll}
\hline\noalign{\smallskip}  
$\Ha$        & $m_1$ & $n$ & $s$ & $f_{\omega}$ & $f_M$   & $f_W$  & $f_{\tau}$ & $f_{\omega_{\text{min}}}$ & $f_{\text{b}}$ &$m_2$ & $f_2$\\
\noalign{\smallskip}\hline\noalign{\smallskip}
5.00         & 2     & 0   & 2   & 42.12     &  2.12  & 21.06 & 2.12    &  42.12              & 2.12       & 2    & 44.24\\
$5.00^{(1)}$  & 2     & 1   & 1   & 42.19     &  4.19  & 12.64 & 2.10    &  38.00              & 19.00      & 1    & 23.89\\
$5.00^{(2)}$  & 2     & 1   & 1   & 42.06     &  4.45  & 12.74 & 2.28    &  37.50              & 18.75      & 1    & 23.31\\
3.5          & 3     & 2   & 1   & 66.05     &  42.86 & 35.91 & 14.29   &  23.19              & 7.73       & 1    & 36.30 \\
3.25         & 3     & 0   & 3   & 66.32     &  22.52 & 22.11 & 22.52   &  66.32              & 22.52      & 3    & 88.85 \\
9.8          & 4     & 1   & 2   & 86.58     &  0.26  & 10.95 & 0.13    &  86.07              & 43.16      & 2    & 43.42\\
\noalign{\smallskip}\hline
\end{tabular}}
\end{center}
\end{table}

Rand's~\cite{Ran82} parameters correspond to the integers $m_1$, $n$
and $s$ (see Eq.~\ref{eq:mod_trav}) and the frequencies
$f_{\omega}=m_1\omega/2\pi$, $f_M=1/\tau_{\text{min}}$ and
$f_W=s(f_{\omega}/m_1+nf_M)/m_1$ (from Eq.~\ref{eq:freq_rand}) whereas
Coughling \& Marcus~\cite{CoMa92} parameters are the integers $m_1$
and $m_2$ and the frequencies $f_{\omega}$ and
$f_2=m_2f_{\omega}/m_1+f_{\tau}$. The results of the classification
are summarised in Table~\ref{table:rand} for the 6 different types of
MRW found at low magnetic forcing corresponding to the radial jet
instability.

We note that with the exception of the MRW$_1^2$, the parent branch of
RW and the corresponding Floquet mode at the bifurcation is
known~\cite{GaSt18} and thus can be used to verify the results
obtained from DNS. For MRW$_1^4$ and MRW$_2^4$ the Laskar frequency
closest to that obtained theoretically at the bifurcation point (see
Table~\ref{table:bif_stab} and Sec.~\ref{sec:eq_waves}) is
$f_{\text{b}}=m_2f_{\omega}/m_1-f_{\tau}$, with $f_{\tau}=1/\tau$. In
contrast, for the MRW$_2^2$ and MRW$_3^3$ we obtain
$f_{\text{b}}=f_{\tau}$. For the MRW$_1^2$ at $\Ha=5$, the origin of
which (bifurcation point) is unknown, we assume
$f_{\text{b}}=m_2f_{\omega}/m_1-f_{\tau}$ as in the case of MRW with
$m_2\ne m_1$.

\begin{figure}
  \includegraphics[scale=0.85]{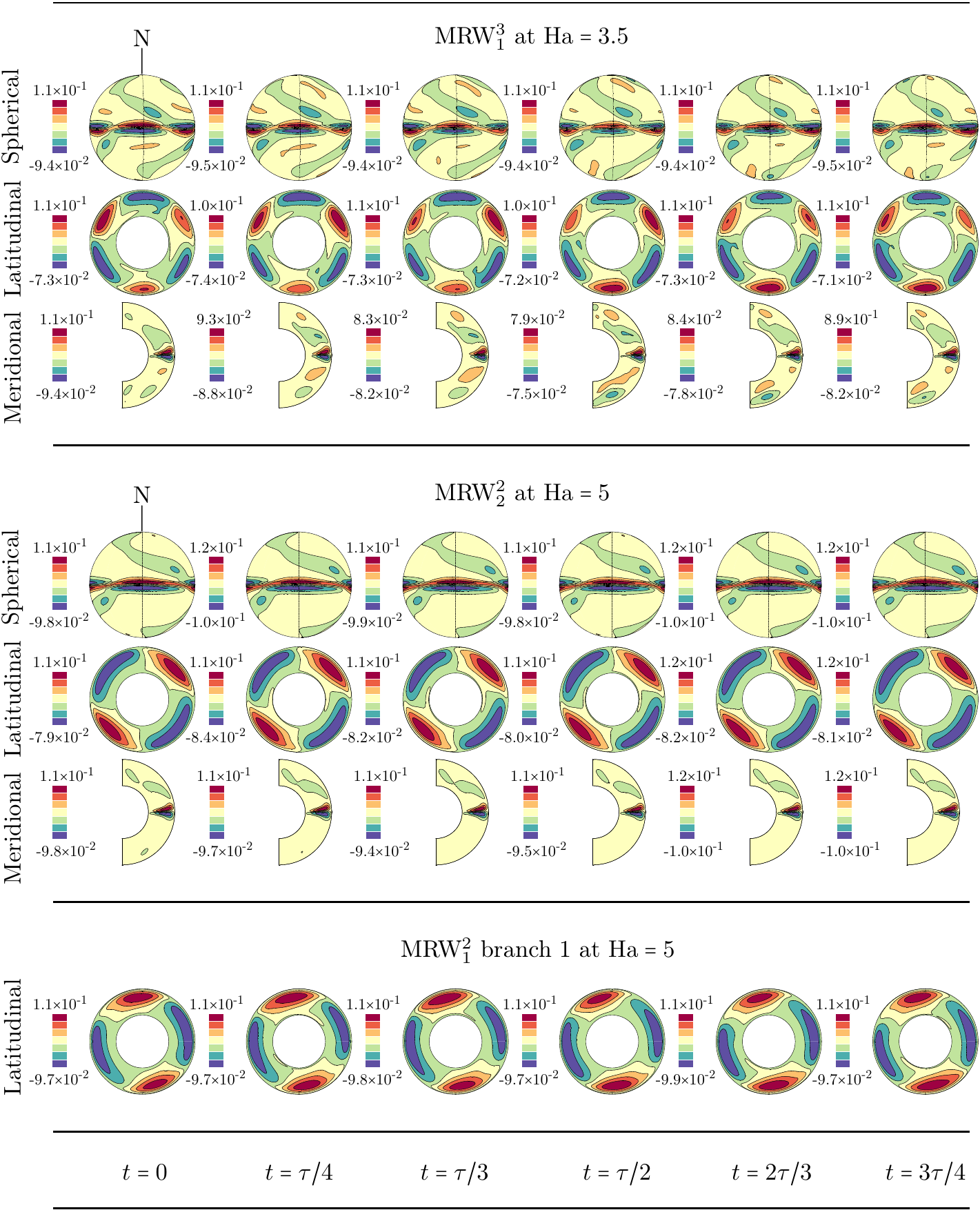}
\caption{Contour plots of the nonaxisymmetric ($m>0$) radial velocity
  on spherical, colatitudinal and meridional cross-sections (from 1st
  to 3rd row) over a period $\tau$ for the MRW$_1^3$ at $\Ha=3.5$. The
  same sections for the MRW$_2^2$ at $\Ha=5$ are from 4th to 6th
  rows. Last row correspond to the colatitude section of the MRW$_1^2$
  lying on branch 1 at $\Ha=5$.  The main azimuthal drift
  corresponding to $\omega$ has been removed. The cross-sections at
  $t=0$ are taken through a maximum of $v_r$. The position of
  colatitudinal and meridional cuts are marked with a vertical line on
  the spherical sections which are displayed from a 0$^\circ$ latidude
  point of view. }
\label{fig:mrw_noaxi}   
\end{figure}

\subsection{Flow patterns for the MRW}

Figure~\ref{fig:mrw_noaxi} visualises the flow patterns and the
spatio-temporal symmetries for three examples of MRW. Specifically,
the nonaxisymmetric radial velocity is displayed at time instants
$t=0,\tau/4,\tau/3,\tau/2,2\tau/3,3\tau/4$ to capture the evolution
over a modulation period $\tau$. The main azimuthal drift
corresponding to $\omega$ has been removed by integrating the system
of Eq.~(\ref{eq:wave_eq}). Spherical, colatitudinal and meridional
cross-sections through a relative maximum of $v_r$ provide a global
view of the flow structure. Supplementary movies displaying the time
evolution over a modulation period, in the frame rotating with the
corresponding frequency (Eq.~(\ref{eq:wave_eq})), are provided for the
colatitudinal sections of each MRW (2nd, 4rth and 7th rows) in
Fig.~\ref{fig:mrw_noaxi}. MRW are strongly equatorially asymmetric in
the form of a radial jet near the equatorial plane as described for RW
corresponding to the radial jet instability~\cite{GaSt18}. However,
for MRW fluid motions start to be noticeable within the cylinder
parallel to the rotation axis and tangent to the inner sphere (see
meridional sections). For the MRW$_1^3$ at $\Ha=3.5$ the patterns at
$t=0$ and $t=\tau/3$ are the same but differ by an azimuthal rotation
of $4\pi/3$ (see colatitudinal sections). Then, $\tau/3$ is the
minimal period and the MRW is of class $(3,2,1)$ according to Rand's
description~\cite{Ran82}. For the MRW$_2^2$ at $\Ha=5$ the patterns at
all time instants are different and thus $\tau$ is the minimal period
and the MRW is of class $(2,0,2)$. Finally, for the MRW$_1^2$ at
$\Ha=5$ the patterns at $t=0$ and $t=\tau/2$ are the same but differ
by an azimuthal rotation of $\pi$. Then, $\tau/2$ is the minimal
period and the MRW is of class $(2,1,1)$.

\section{Dominant azimuthal mode analysis}
\label{sec:m_dom}

This section is devoted to a detailed description of the main
azimuthal modes involved in four examples of MRW. They are MRW$_1^3$
at $\Ha=3.5$, MRW$_2^2$ at $\Ha=5$, and two MRW$_1^2$ on branch 1 and
2, both at $\Ha=5$. The later choice makes sense as experimental data
at $\Ha=5$ is already available~\cite{KKSS17}. For each example of MRW
we select three sets of positive azimuthal wave numbers in its
spherical harmonics expansion
(Eqs.~(\ref{eq:serie_psi}-\ref{eq:serie_phi})) containing more than
$96\%$ of the nonaxisymmetric ($m>0$) kinetic energy. Apart from being
the most energetic modes, the choice of the sets is also motivated to
reflect the azimuthal symmetry of the underlying RW and that of the
perturbation (Floquet mode) giving rise to the modulation. Indeed, the
MRW can then be reasonably approximated by the sum of the three sets
of modes rather than by the full expansion of
Eqs.~(\ref{eq:serie_psi}-\ref{eq:serie_phi}). From a time integration
of a MRW, with the main azimuthal drift removed (i.\,e. integrating
the system Eq.~(\ref{eq:wave_eq})), we can follow individually the
evolution of each set (the sum of their modes) over a modulation
period $\tau$. This provides useful information about the location of
maximum flow velocities for specific azimuthal symmetries and hence is
important from the experimental point of view, as flow measurement
probes in the HEDGEHOG experiment can be located adequately to capture
certain azimuthal symmetries.

For each set of modes we consider individually their equatorially
symmetric as well as antisymmetric component to study in detail the
equatorial symmetry. MRW are dominated by the antisymmetric
contribution as is characteristic for radial jet instabilities at low
magnetic field~\cite{Hol09}. Measurement probes located symmetrically
with respect to the equator help to infer the equatorially symmetric
(ES) as well as antisymmetric (EA) part of the flow in experimental
devices~\cite{KKSS17}. In addition, any measurement in the equatorial
plane refers only to the ES part of the flow (as the EA part of any
field is zero at the equator). The dominant azimuthal wave number of
the ES flow $m^{\text{s}}_{\text{max}}$ may differ from the dominant
azimuthal wave number ($m_{\text{max}}$) of the total solution which,
indeed is the case for our examples.  We have found
$m^{\text{s}}_{\text{max}}=2m_{\text{max}}$ for all the classes of
MRW. Then, the time series of a scalar field at a point
$(r,\pi/2,\varphi)$ (equator) will have half the period than that at a
point $(r,\theta,\varphi)$, $\theta\ne\pi/2$ (off the equator), which
is verified in our simulations.

\begin{figure}
  \includegraphics[scale=0.85]{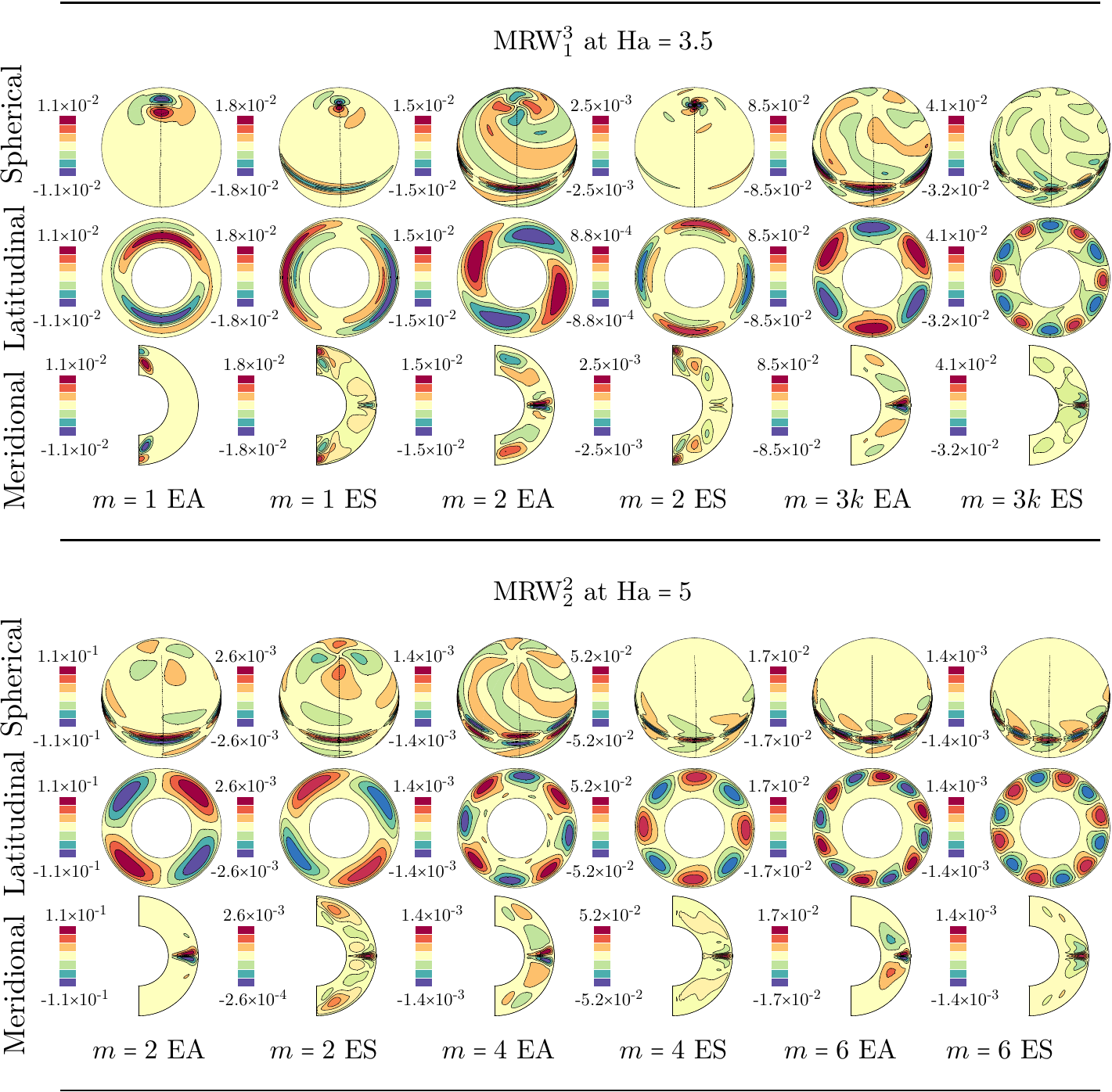}
\caption{Most energetic modes with $m=1$, $m=2$ and $m=3k$
  $k\in\mathbb{Z}$, equatorially antisymmetric (EA) or symmetric modes
  (ES) of the MRW$^3_1$ at $\Ha=3.5$ (top 3 rows) and modes $m=2,4,6$
  of the MRW$_2^2$ at $\Ha=5$ (bottom 3 rows). Cross-sections are as
  Fig.~\ref{fig:mrw_noaxi} but the spherical section point of view is
  at 45$^\circ$ latitude.}
\label{fig:mrw_cpt_mod1}   
\end{figure}

\subsection{MRW$^3_1$ at $\Ha=3.5$}

For the MRW$^3_1$ at $\Ha=3.5$ we consider three set of modes $m=1$,
$m=2$ and $m=3k$, $k\in\mathbb{Z}$ (containing all the modes which are
multiples of $3$). Their kinetic energy content is
$K_{m=1}/K_{m>0}=0.03$, $K_{m=2}/K_{m>0}=0.05$ and
$K_{m=3k}/K_{m>0}=0.91$, where $K_{m}$ is the time and volume averaged
kinetic energy contained in each set of modes and $K_{m>0}$ is the
nonaxisymmetric kinetic energy. Figure~\ref{fig:mrw_cpt_mod1} displays
a snapshot of the nonaxisymmetric radial velocity in spherical,
colatitudinal and meridional cross-sections similar to
Fig.~\ref{fig:mrw_noaxi}. Each column represents the ES or EA part of
each set of modes extracted from the spherical harmonics expansion of
the total flow. Supplementary movies displaying the time evolution
over a modulation period, in the frame rotating with the corresponding
frequency (Eq.~(\ref{eq:wave_eq})), are provided for each
colatitudinal section in 2nd and 5th rows of
Fig.~\ref{fig:mrw_cpt_mod1}. We recall that the corresponding time
evolution of the nonaxisymmetric part of the flow (also provided as
supplementary movie) will be accurately approximated by the sum of the
flow components shown in Fig.~\ref{fig:mrw_cpt_mod1}. The inspection
of the latter figure and its associated movies provide relevant
information summarised in the following (also in
Table~\ref{table:mode_desc}):
\begin{itemize}
\item The set $m=3k$ is clearly associated with the underlying RW. The
  EA part (5th column of Fig.~\ref{fig:mrw_cpt_mod1}) of this set of
  modes is dominant as the maximum radial velocity is more than twice
  that of the ES part (6th column of Fig.~\ref{fig:mrw_cpt_mod1}). In
  addition, the EA part is steady as is the case of a RW. The ES part
  is not azimuthally drifting, but changing its shape with a period
  $\tau/2$ associated to the modulation. The azimuthal symmetry of the
  EA ($m=3$) and ES ($m=6$) part is different (see colatitude
  sections) and thus measured velocities for the EA and ES part
  exhibit different periods (differing by a factor of two).
\item The set $m=1$ only contains a single azimuthal wave number and
  is related with the Floquet multiplier giving rise to this branch of
  MRW. The radial velocity patterns (1st and 2nd columns) resemble
  quite well those of the Floquet multiplier given in Fig.~7
  of~\cite{GaSt18} and evidence that polar fluid motions appear
  noticeable for the MRW$_1^3$ but are very weak for the RW$_3^3$
  (compare meridional sections of 1st/2nd columns with 5th/6th
  columns). The time evolution of the $m=1$ EA part corresponds to a
  polar vortex with a purely retrograde azimuthal rotation (i.\,e. the
  pattern remains the same) of period $\tau/2$ whilst in the ES case,
  apart from the azimuthal drift (in the prograde way), there is a
  very weak change of pattern with period $\tau$.
\item The flow patterns of the single mode set $m=2$ are related with
  those of the $m=1$ as well as $m=3k$ sets because of the nonlinear
  coupling between the modes. Fluid motions develop in the polar as
  well as the equatorial region and the time dependence is similar to
  that of the $m=1$ set. The EA part rotates azimuthally (retrograde)
  without changing shape and is $\tau$ periodic whilst the ES part is
  prograde but changing shape and the period is $\tau/2$.
\end{itemize}
We note that for the three sets of modes the observed change of shape
is very weak or absent.  From this analysis we conclude the functional
form of a MRW in the MSC system is more complicated than that given in
Eq.~(8) (or Eq.~(12)) of~\cite{CoMa92} in the context of
Taylor-Couette flow. In particular, time dependence is not only
associated to the $\varphi$ (azimuthal) coordinate as
in~\cite{CoMa92}, but also to the radial and colatitudinal coordinates
as is reflected by the change of shape of the ES part of the flow
(Table~\ref{table:mode_desc}). However, their functional form remains
valid in the case of the EA part of the flow because no shape change
is observed in the rotating frame and the patterns are only
azimuthally drifting. In addition, because the EA part of the flow is
larger than the ES part and the shape changes associated with this
part are quite weak (see supplementary movies), its volume averaged
properties exhibit very small modulations as reproduced in the time
series of Fig.~\ref{fig:ts_m3}(b,c).

\begin{table} 
\caption{Summary of the time evolution (in the frame of reference
  rotating with frequency $\omega$) of each set of dominant modes for
  the MRW$_1^3$ at $\Ha=3.5$, the MRW$_2^2$ at $\Ha=5$, and the
  MRW$_1^2$ on branch 1 and 2, both at $\Ha=5$. In all the modes the
  observed change of shape is very weak or absent and the pattern is
  azimuthally rotating or fixed. In some cases (marked with *) the
  azimuthal drift is nonuniform. Polar or equatorial position of the
  maximum of the radial velocity of each mode is also stated. The time
  evolution of each mode (each row in the table) over a period $\tau$
  is displayed in the supplementary movies for the MRW$^3_1$ and
  MRW$^2_2$.}
\label{table:mode_desc}
\begin{center}
  MRW$^3_1$ at $\Ha=3.5$\\[5.pt]
\scalebox{0.9}{  
\begin{tabular}{lllllll}
\hline\noalign{\smallskip}  
$m$  & Eq. Sym. & $\max v_r$ & period   & drift       & shape change & Position $\max v_r$ \\
\noalign{\smallskip}\hline\noalign{\smallskip}
$1$  & -1       & $0.011$     & $\tau/2$ & retrograde & 0            & Polar \\
$1$  &  1       & $0.018$     & $\tau$   & prograde   & 1 (weak)     & Polar \\
$2$  & -1       & $0.015$     & $\tau$   & retrograde & 0            & Equatorial \\
$2$  &  1       & $0.0025$    & $\tau/2$ & prograde   & 1            & Polar\\
$3k$ & -1       & $0.085$     & $-$      &  none      & 0            & Equatorial \\
$3k$ &  1       & $0.041$     & $\tau/2$ &  none      & 1            & Equatorial \\
\noalign{\smallskip}\hline
\end{tabular}}\\[5.pt]
MRW$^2_2$ at $\Ha=5$\\[5.pt]
\scalebox{0.9}{  
\begin{tabular}{lllllll}
\hline\noalign{\smallskip}  
$m$  & Eq. Sym. & $\max v_r$ & period   & drift       & shape change & Position $\max v_r$ \\
\noalign{\smallskip}\hline\noalign{\smallskip}
$2$  & -1       & $0.11$    & $\tau/2$ & none       & 1 (weak)     & Equatorial \\
$2$  &  1       & $0.0026$  & $\tau$   & prograde*  & 1            & Equatorial \\
$4$  & -1       & $0.0014$  & $\tau$   & prograde*  & 1            & Equatorial \\
$4$  &  1       & $0.052$   & $\tau/2$ & none       & 1 (weak)     & Equatorial \\
$6$  & -1       & $0.017$   & $\tau/2$ & none       & 1            & Equatorial \\
$6$  &  1       & $0.0014$  & $\tau$   & prograde*  & 1            & Equatorial \\
\noalign{\smallskip}\hline
\end{tabular}}\\[5.pt]
MRW$^2_1$ (branch 1) at $\Ha=5$\\[5.pt]
\scalebox{0.9}{  
\begin{tabular}{lllllll}
\hline\noalign{\smallskip}  
$m$  & Eq. Sym. & $\max v_r$ & period   & drift       & shape change & Position $\max v_r$ \\
\noalign{\smallskip}\hline\noalign{\smallskip}
$1$  & -1       & $0.0048$     & $\tau$   & prograde*   & 1            & Equatorial \\
$1$  &  1       & $0.0089$     & $\tau$   & retrograde* & 1  (weak)    & Equatorial \\
$2k$ & -1       & $0.094$      & $\tau/2$ & none        & 1            & Equatorial \\
$2k$ &  1       & $0.054$      & $\tau/2$ & none        & 1            & Equatorial \\
$3$  & -1       & $0.0075$     & $\tau$   & retrograde* & 1            & Equatorial \\
$3$  &  1       & $0.0087$     & $\tau$   & prograde*   & 1            & Equatorial \\
\noalign{\smallskip}\hline
\end{tabular}}\\[5.pt]
MRW$^2_1$ (branch 2) at $\Ha=5$\\[5.pt]
\scalebox{0.9}{  
\begin{tabular}{lllllll}
\hline\noalign{\smallskip}  
$m$  & Eq. Sym. & $\max v_r$ & period   & drift       & shape change & Position $\max v_r$ \\
\noalign{\smallskip}\hline\noalign{\smallskip}
$1$  & -1       & $0.0033$    & $\tau$   & prograde*   & 1            & Equatorial \\
$1$  &  1       & $0.017$     & $\tau$   & retrograde* & 1 (weak)     & Equatorial \\
$2k$ & -1       & $0.093$     & $\tau/2$ & none        & 1            & Equatorial \\
$2k$ &  1       & $0.05$      & $\tau/2$ & none        & 1            & Equatorial \\
$3$  & -1       & $0.015$     & $\tau$   & retrograde* & 1 (weak)     & Equatorial \\
$3$  &  1       & $0.0091$    & $\tau$   & prograde*   & 1            & Equatorial \\
\noalign{\smallskip}\hline
\end{tabular}}
\end{center}
\end{table}

\begin{figure*}
  \includegraphics[scale=0.85]{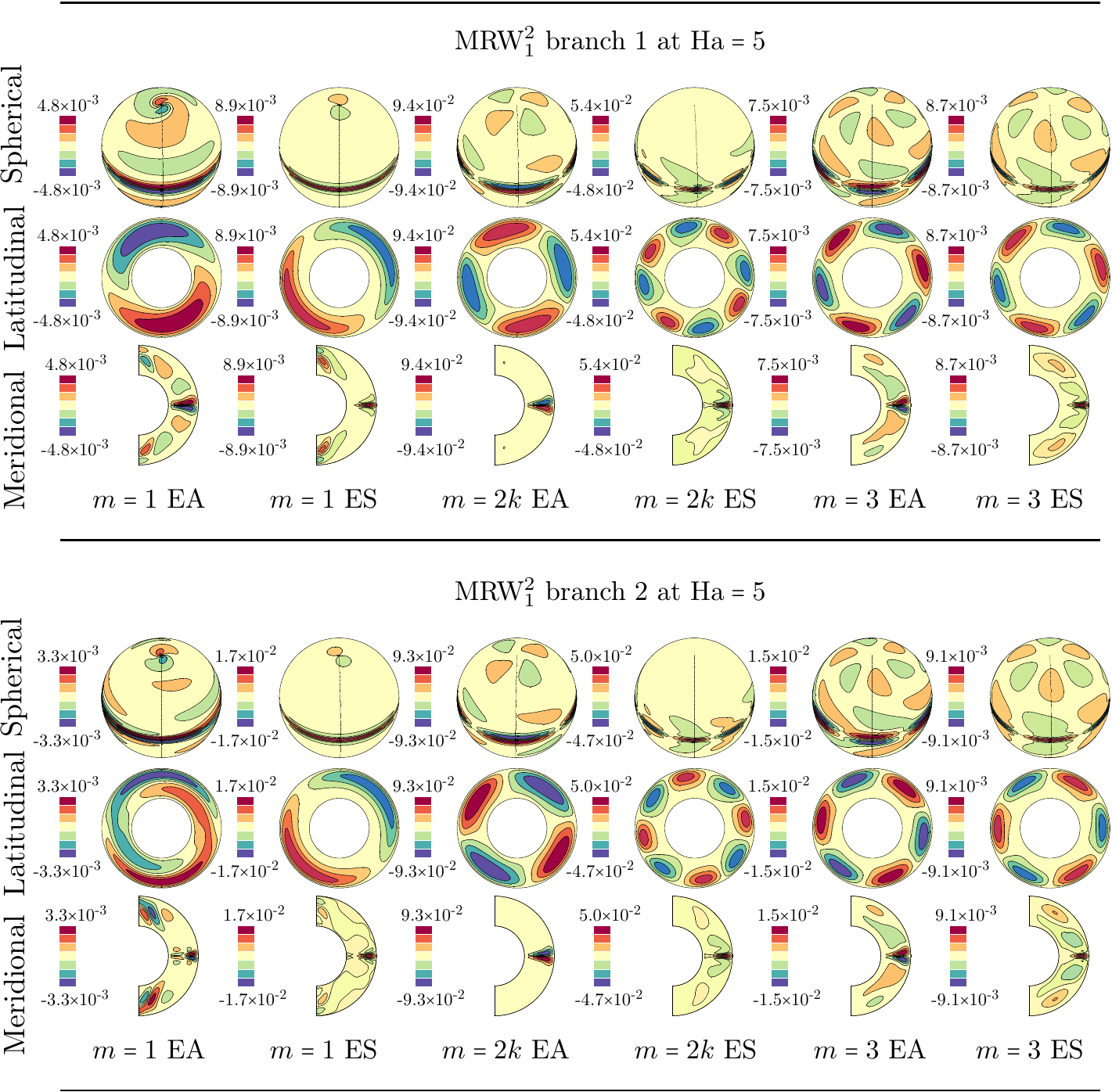}
\caption{Most energetic $m=1,2k,3$ equatorially antisymmetric (EA) or
  symmetric modes (ES) of the MRW$^2_1$ at $\Ha=5$ on branch 1 (top 3
  rows) and of the MRW$_1^2$ on branch 2 at $\Ha=5$ (bottom 3
  rows). Cross-sections are as Fig.~\ref{fig:mrw_noaxi} but the
  spherical section point of view is at 45$^\circ$ latitude.}
\label{fig:mrw_cpt_mod2}   
\end{figure*}

\subsection{MRW$^2_2$ at $\Ha=5$}

The three sets of modes for the MRW$^2_2$ at $\Ha=5$ are the most
energetic nonaxisymmetric modes $m=2$, $m=4$ and $m=6$. Their kinetic
energy ratios are $K_{m=2}/K_{m>0}=0.86$, $K_{m=4}/K_{m>0}=0.10$ and
$K_{m=6}/K_{m>0}=0.03$.  The outcomes of the mode analysis for the
MRW$_2^2$ are presented in Fig.~\ref{fig:mrw_cpt_mod1} (three bottom
rows) and Table~\ref{table:mode_desc} (also supplementary movies). The
results are different for the case of MRW$_1^3$. Now the time
evolution (recall Eq.~(\ref{eq:wave_eq})) for the three sets, their ES
as well as EA part, exhibit a change of the pattern while it is
drifting or steady in the azimuthal direction. The significant change
of shape of the modes results in noticeable oscillations of the volume
averaged properties (see Fig.~\ref{fig:bif_diagr}(b) at $\Ha=5$ on the
MRW$_2^2$ branch). As for the MRW$_1^3$ the ES and EA part of each
mode have different periods $\tau/2$ and $\tau$. We note that the
azimuthal drift is always prograde but develops in a nonuniform
fashion and that the dominant azimuthal wave number of the EA part of
the flow is $m=2$ while that of the ES part is doubled as is the case
for the MRW$_1^3$. Although the sets of modes (columns in
Fig.~\ref{fig:mrw_cpt_mod1}) have the position of the maximum
nonaxisymmetric radial velocity in the equatorial region, polar fluid
motions are noticeable in the ES $m=2$ as well as EA $m=4$ modes (also
slightly the $m=6$ ES mode, see meridional sections). The time
dependence of the EA (ES) component of the $m=4$ mode is qualitatively
similar to the ES (EA) component of the $m=2,6$ modes, exhibiting a
stronger change of shape.  This may be an indication that these flow
components are strongly correlated with the corresponding Floquet mode
giving rise to the MRW$_2^2$ branch.

\subsection{MRW$^2_1$ at $\Ha=5$: branch 1 and 2}

Finally, the sets for the MRW$^2_1$ on branch 1 and 2 at $\Ha=5$ are
the modes $m=1$, $m=2k$ and $m=3$ which have $K_{m=1}/K_{m>0}=0.007$,
$K_{m=2k}/K_{m>0}=0.94$ and $K_{m=3}/K_{m>0}=0.032$ in the case of
branch 1 and $K_{m=1}/K_{m>0}=0.017$, $K_{m=2k}/K_{m>0}=0.87$ and
$K_{m=3}/K_{m>0}=0.080$ for the MRW$^2_1$ on branch 2. The time
evolution of all the sets of modes in both branches is very
similar. The patterns for each set are shown in the columns of
Fig.~\ref{fig:mrw_cpt_mod2} and their type of time dependence is
outlined in Table~\ref{table:mode_desc}. As happened for the other MRW
the dominant azimuthal wave number of the EA part of the flow is $m=2$
while that of the ES part is doubled to $m=4$. In addition, noticeable
polar fluid motions are present in the $m=1$ mode and the $m=2k$ mode
does not exhibit any azimuthal drift. This gives more support for the
hypothesis that MRW$^2_1$ bifurcate from a RW branch with $m=2$
azimuthal symmetry and that their Floquet mode has azimuthal symmetry
$m=1$. Moreover, the most noticeable difference between the MRW$^2_1$
on both branches is seen in the $m=1$ and $m=3$ mode while the $m=2k$
modes behave quite similarly ($\omega$ and $\tau$ are quite similar as
well, see Table.~\ref{table:rand}). This may be an indication that
both branches bifurcate at different points of the same branch of RW
with azimuthal symmetry $m=2$. To fully understand the origin of the
two branches of MRW$^2_1$ a continuation method for MRW as
in~\cite{GNS16} must be used.

\section{Summary}
\label{sec:sum}

Modulated rotating waves (MRW) in the magnetized shperical Couette
system were computed for the radial jet instability~\cite{Hol09} and
few for the return flow~\cite{HoSk01} instability. The former for
Hartmann numbers $\Ha<10$ and the later very close to $\Ha=30$, both
for fixed Reynolds number $\Ree=10^3$. These values correspond to
experimental runs of the HEDGEHOG experiment already
obtained~\cite{KKSS17} and fall in the interval where branches of
rotating waves (RW) with $m=2,3,4$-fold azimuthal symmetry have been
recently obtained~\cite{GaSt18}. MRW are obtained by means of direct
numerical simulations (DNS) from initial conditions given by the
stability (Floquet) analysis of RW performed in~\cite{GaSt18} which
provides estimations for the time scales and symmetries of bifurcated
MRW. The main findings of the present study are summarized in the
following.

For the equatorially asymmetric radial jet instability ($\Ha\sim 5$)
several types of MRW, and even 3 frequency time dependent flows, are
described and their regions of stability determined. Several large
intervals of multistability of MRW are found. This parameter regime is
a good candidate for searching MRW in the HEDGEHOG experiment. In
contrast, in case of the equatorially symmetric return flow
instability ($\Ha\sim 30$) MRW are very rare and their regions of
stability very narrow. Almost only rotating waves (RW) described
in~\cite{GaSt18} can be found because of the moderate value of
$\Ree$. This parameter regime is then a good candidate for searching
RW in the HEDGEHOG experiment.  Broader stability intervals for MRW
corresponding to the return flow instability may appear with
increasing $\Ree$ as shown in~\cite{Hol09} for the aspect ratio
$\chi=0.33$.

From the RW with three-fold azimuthal symmetry, bifurcated from the
base axisymmetric state at $\Ha_{\text{c}}=12.2$, the typical
Ruelle-Takens scenario, giving rise to $m=1$-fold azimuthal symmetric
MRW and 3 frequency radial jet time dependent flows, has been found by
decreasing the Hartmann number. The flows belonging to this scenario
have their kinetic energy mostly contained in the $m_{\text{max}}=3$
azimuthal wave number matching the symmetry of the parent RW. However,
these MRW do not exhibit noticeable oscillations of volume averaged
physical properties, as is commonly found~\cite{Hol09,GSDN15}, which
may lead a wrong classification as RW if only volume-averaged
properties are investigated.



We follow the theoretical studies of~\cite{Ran82} and~\cite{CoMa92}
and provide a description of the different types of MRW found in terms
of their classification. In particular, the spatio-temporal symmetries
of the MRW~\cite{Ran82} and their connection with RW and Floquet
modes~\cite{CoMa92} are rigorously stated. The figures and
supplementary movies along a modulation period for a representative
set of MRW provide an easy visualisation of the theoretical
classification.

By investigating the evolution of the most energetic azimuthal wave
numbers in the spherical harmonics expansion of a MRW, a careful
analysis of the spatio-temporal symmetries of the different azimuthal
modes (equatorially symmetric or antisymmetric) for a selected types
of MRW corresponding to the radial jet instability is performed. The
study of MRW in terms of its dominant modes with $m=1,2,3$-fold fixed
azimuthal symmetry allows us to determine the regions (equatorial or
polar) within the shell with maximum signal (i.\,e. large flow
velocities) for each class of azimuthal symmetry in order to constrain
the measurement set-up in the HEDGEHOG experiment.

A characteristic difference between MRW and their corresponding parent
RW is that the onset of modulation is associated with the appearance
of noticeable flow velocities in the polar regions. A careful
inspection of the equatorial symmetry evidences that the dominant
azimuthal wave number of the flow $m_{\text{max}}$ corresponds to that
of the antisymmetric flow whilst it is doubled in the case of
symmetric flow. This is relevant for future runs of the HEDGEHOG
experiment as any measurement in the equatorial plane refers to the
equatorially symmetric part of the flow.

Finally, we give numerical evidence that the time evolution of a
specific single mode $m$, viewed in the frame azimuthally drifting
with the corresponding frequency, can not only be associated with the
azimuthal coordinate as in~\cite{CoMa92}, hence a time dependence on
the radial as well colatitudinal coordinates needs to be
considered. MRW with nearly constant mean properties can be accurately
approximated by nonlinear interaction of their dominant modes, which
are either almost steadily azimuthally rotating or fixed exhibiting a
weak change of the flow topology as the time dependence of their
volume-averaged properties suggests. MRW with oscillating mean
properties have some dominant modes, which are changing their pattern
while also are azimuthally rotating or fixed. For these MRW the
azimuthal drift seen in some azimuthal modes takes place in nonuniform
fashion.

\section*{Acknowledgements}

F. Garcia kindly acknowledges the Alexander von Humboldt Foundation
for its financial support.

\bibliographystyle{spmpsci}      


\end{document}